\documentclass[twocolumn,prl,showpacs,superscriptaddress,floatfix]{revtex4}
\usepackage{graphicx}
\usepackage{dcolumn}
\usepackage{amsmath}
\usepackage{amssymb}
\usepackage{latexsym}
\usepackage{bm}

\begin{document}

\title{Particle Condensation in Pair Exclusion Process}

\author{Sang-Woo Kim}
\affiliation{Department of Physics, University of Seoul, Seoul 130-743, Korea}
\author{Joongul Lee}
\affiliation{Department of Mathematics Education, Hongik University, Seoul 121-791, Korea}
\author{Jae Dong Noh}
\affiliation{Department of Physics, University of Seoul, Seoul 130-743, Korea}

\date{\today}
\begin{abstract}
Condensation is characterized with a single macroscopic condensate 
whose mass is proportional to a system size $N$. 
We demonstrate how important particle interactions are in condensation
phenomena. We study a modified version of the zero-range process 
by including a pair exclusion. Each particle is
associated with its own partner, and particles of a pair are forbidden to 
stay at the same site. The pair exclusion is weak in that a particle
interacts with only a single one among all others. It turns out that
such a weak interaction changes the nature of condensation
drastically. There appear a number of mesoscopic condensates: 
the mass of a condensate scales as $m_{\rm con}\sim N^{1/2}$ and
the number of condensates scales as $N_{\rm con} \sim N^{1/2}$ with a
logarithmic correction. These results are derived analytically through a
mapping to a solvable model under a certain assumption, and confirmed
numerically.
\end{abstract}
\maketitle

Condensation occurs in various systems ranging from 
equilibrium ideal Bose gases to nonequilibrium lattice gases~(see 
Ref.~\cite{Evans05} and references therein).
In a condensed phase, a finite fraction of particles constitute a
macroscopic condensate at a single microscopic state which is either favored 
or selected spontaneously. 
Since particles are highly populated in a condensate, 
particle interaction can play an important role~\cite{Evans06,Waclaw09}.
In this paper, we investigate robustness of a macroscopic condensate 
against a weak particle repulsion. 

Our work is motivated from a recent work on 
evolving networks which exhibit a condensation of edges at some vertices 
called hubs~\cite{Noh08_09}. The edge condensation becomes analogous 
to a particle condensation if one interprets vertices and edges as
lattice sites and particles, respectively. 
Interestingly, the edge condensation shows a distinct feature 
that there are a large number of mesoscopic condensates, the origin of which
has not been understood yet. 
We notice that networks are subject to the constraint which forbids 
self-loops. Although it has a negligible effect on network 
properties in most cases, it may be responsible for the
unusual condensation. 
We will investigate a lattice gas which is equipped with a so-called pair 
exclusion mimicking the self-loop constraint.

We study the zero-range process~(ZRP)~\cite{Spitzer70}
which is useful for the study of condensation~\cite{Jeon00,Evans05}.
Suppose that there are $M=\rho N$ particles 
in an arbitrary lattice of $N$ sites with particle density $\rho$. 
A lattice site $i=1,\ldots,N$ can accommodate many particles, and its
occupation number $m_i$ will be called a mass.
Each site $i$ emits a particle at a mass-dependent rate $u(m_i)$. 
The canonical form is given by
\begin{equation}\label{canonical_u}
u(m) = 1+\frac{b}{m}  
\end{equation}
with a control parameter $b$.
An emitted particle then tries to hop to a site $j$
selected with the probability given by a stochastic hopping matrix $\bm{T} = 
\{T_{ji}\}$ which reflects
the structure of the underlying lattice and the presence of any bias in
dynamics.  
We now introduce the pair exclusion~(PE): 
Each particle is associated with a unique partner 
so that there are $M/2$ distinct pairs in total with even $M$. 
A hopping will be rejected if it would result in occupation of a pair 
at the same site. 
The resulting model will be referred to as a pair exclusion process~(PEP) in
short. A particle interacts with its partner only. 
Such a weak interaction changes the nature
of condensation drastically as we shall see below.

In the ZRP without PE, the particle hopping rate depends only on the mass at
departing sites. This property yields that
the mass distribution in the stationary state is factorized
into a product form~\cite{Evans05}.
Particle hopping rates in the PEP, however, depend on the whole 
distribution of particle species, which makes it difficult to treat 
the problem analytically. Hence, we adopt a series of approximations. 

We first assume that the particle species distribution is purely random.
Then, a configuration of the system can be specified 
with a mass distribution $\bm{m} = (m_1,\ldots,m_N)$, 
and the particle hopping rate from a site $i$ to $j$ 
is given by~\cite{unpub}
\begin{equation}\label{Wt_ji}
\widetilde{W}_{ji}(m_j,m_i) =  \tilde{v}(m_j,m_i) \ T_{ji} \ u(m_i)  ,
\end{equation}
with 
\begin{equation}\label{pep_prob} 
\tilde{v}(m,m') = 1- \lambda \frac{\sum_{k=0}^{\min (m,m')} k\binom{m'}{k}\binom{M/2-m'}{m-k}2^{m-k}}{\sum_{k=0}^{\min (m,m')} m'\binom{m'}{k}\binom{M/2-m'}{m-k}2^{m-k}}.
\end{equation}
The factor $\tilde{v}(m,m')$ accounts for the probability 
that a particle selected from a site of mass $m'$ does not find 
its partner at another site of mass $m$.
For convenience, we insert a parameter $\lambda=1$. 
One can turn off the PE by choosing $\lambda=0$. With $0<\lambda<1$, 
one can deal with a pair repulsion that forbids a pair occupation 
with probability $\lambda$.

For large $M\gg m$ and $m'$, one can expand $\tilde{v}(m,m')$ as 
$\tilde{v}(m,m') = v(m) + \mathcal{O}((m/M)^2)$ with 
\begin{equation}\label{def_vt}
{v}(m) = 1 - \lambda \frac{m}{M} = 1-\lambda \frac{m}{\rho N} .
\end{equation}
We make an additional approximation by ignoring the $\mathcal{O}((m/M)^2)$ term.
Then, the hopping rate is given by 
\begin{equation}\label{W_ji}
W_{ji}(m_j,m_i) = v(m_j) \ T_{ji} \ u(m_i)  \ .
\end{equation}

The hopping rate depends on the mass at both departing and 
arriving sites. It is interesting to show whether the stationary state of
such a system is also given by a factorized product state as the ZRP. 
The factorizability condition for the general hopping rate has been 
investigated recently in an one-dimensional lattice or a complete 
graph~\cite{Luck07}. 
The hopping rate in Eq.~(\ref{W_ji}) is specifically given by the product 
of $u(m)$ and $v(m)$.  We will show that such a
system has a factorized stationary state for any functions $u(m)$ and $v(m)$
when the hopping matrix $\bm{T}$ satisfies the {\em detailed balance} condition.

Let $P(\bm{m};t)$ be the probability of the system being in a configuration
$\bm{m} = (m_1,\ldots,m_N)$ at time $t$. The
time evolution follows the master equation
\begin{eqnarray}\label{master_eq}
\frac{\partial P(\bm{m};t)}{\partial t} &=&  \sum_{i,j=1}^N
\left\{ - W_{ji}(m_j,m_i) P(\bm{m};t) \right. \nonumber \\
&+& \left. W_{ji}(m_j-1,m_i+1) P(\bm{m}'_{ji};t)\right\} ,
\end{eqnarray}
where $\bm{m}'_{ji}=(m'_1,\ldots,m'_N)$ is a shorthand notation for the
the mass distribution with $m'_i = m_i +1$, $m'_j = m_j-1$, and 
$m'_k = m_k$ for $k\neq i,j$. 

We seek for the stationary state solution in a product form given by
\begin{equation}\label{P_st}
P_{st}(\bm{m}) = \frac{1}{Z(N,M)} \left[ \prod_{i=1}^N f_i(m_i) \right]
\delta_{M,\sum_{j=1}^N m_j}  ,
\end{equation}
where $\delta_{a,b}$ is the Kronecker-$\delta$ symbol, 
$Z(N,M)$ is a normalization, and $f_i(m)$'s have to be determined. 
Inserting $P_{st}$ into the right-hand side of Eq.~(\ref{master_eq}) and
taking it to zero, we find
that $f_i(m)$ is given by 
\begin{equation}\label{f_sol}
f_i(m) = \prod_{n=1}^m \left[ \omega_i \frac{ v(n-1)}{u(n)} \right]
\end{equation}
for $m>1$ and $f_i(0) = 1$. Here the auxiliary constant 
$\{\omega_i\}$ should satisfy 
\begin{equation}\label{solvability}
\sum_{i,j=1}^N \left( \frac{\omega_i}{\omega_j} T_{ji} - T_{ij} \right) v(m_i) u(m_j) 
=0 \ .
\end{equation}
For arbitrary functions $u(m)$ and $v(m)$, it requires 
\begin{equation}\label{detailed_balance}
T_{ji}\ \omega_i = T_{ij}\ \omega_j 
\end{equation}
for all $i$ and $j$.

The stochastic matrix $\bm{T}$ by itself defines a hopping
dynamics of a single particle. The condition~(\ref{detailed_balance})
is equivalent to the detailed balance condition for the corresponding 
single particle dynamics. 
Hence, the PEP in one-dimensional rings, for example, has a
factorized product state with $\{\omega_i=1\}$ when the hopping is symmetric. 
When there is a bias in hopping, the factorization is not guaranteed.
The role of the hopping
bias is discussed in a target process where the hopping rate depends only
on the mass at arriving sites~\cite{Luck07}.  
We note that the PEP in the solvable case has the same stationary
state as the ZRP with particle emitting rate $\tilde{u}(m) =
u(m)/v(m-1)$.

The detailed balance is a sufficient condition for the 
factorization. 
One may also have a factorized state without the detailed balance
for a specific $u(m)$ and $v(m)$.
The ZRP corresponds
to the specific case with $v(m)=1$.
Using $\sum_{i}T_{ij}=1$, one can show that
the relation~(\ref{solvability}) is fulfilled by $\{\omega_i\}$
satisfying $\sum_{j=1}^N T_{ij} \omega_j = \omega_i$
for all $i$~\cite{Noh05}. The solution $\{\omega_i\}$ is nothing but 
the stationary state probability of a single particle whose dynamics is 
governed by $\bm{T}$, which exists for any stochastic matrix $\bm{T}$.

In order to investigate the effect of the PE on the condensation, 
we consider the PEP with 
the symmetric hopping in an one-dimensional ring or the random hopping in a
complete graph.
In both cases, $\omega_i=1$ and $P_{st}(\bm{m})$
is given by Eq.~(\ref{P_st}) with $f_i(m)=f(m)$ for all $i$, where
\begin{equation}\label{f_prod}
f(m) = \prod_{n=1}^m \left[ \frac{1-\lambda(n-1)/(\rho N)}{1+b/n} \right] .
\end{equation}
For $ m \ll N$, it is given by 
\begin{equation}\label{f_factor}
f(m) \simeq \frac{\Gamma(m+1) \Gamma(b+1)}{\Gamma(m+b+1)}
\exp\left[ -\frac{\lambda m^2}{2\rho N} \right] 
\end{equation}
with the gamma function $\Gamma(s)\equiv\int_{0}^{\infty} x^{s-1}e^{-x}dx$.
The single-site mass distribution function $p(m)$, defined as the probability 
that the mass of a site is $m$, is given by 
\begin{equation}\label{pm_def}
p(m) = f(m) {Z(N-1,M-m)}/{Z(N,M)} .
\end{equation}

The PE imposes an upper cutoff at
$m_{PE}=\mathcal{O}(N^{1/2})$ in the mass distribution. 
The knowledge on the ZRP~($\lambda=0$) allows us to 
guess the phase behavior of the PEP.
When $\lambda=0$, the system is in a fluid phase for $\rho<\rho_c = 1/(b-2)$
and in a condensed phase for $\rho>\rho_c$.
The former is characterized with $p(m)\sim m^{-b}
e^{-m/m_0}$ with a constant $m_0 = \mathcal{O}(1)$, while the latter 
with $p(m)$ displaying a power-law scaling $p(m)\sim m^{-b}$ for bulk sites 
and a peak around $m=(\rho-\rho_c)N$ for a macroscopic condensate.
When the PE is turned on, the fluid phase will not be affected 
because the mass distribution is limited to the region $m \lesssim m_0$
where the PE is negligible~($m_0 \ll m_{PE})$.
On the other hand, the system cannot maintain the
macroscopic condensate in the presence of PE because of the cutoff $m_{PE}$.
The excess mass cannot be absorbed into bulk sites because the mean
occupation number at bulk sites is limited by $\rho_c$.
A possible solution is to build up
$(\rho-\rho_c)N/m_{PE}$ mesoscopic condensates of mass $m \sim m_{PE}$. Hence we
expect that the PE does not modify the phase boundary but 
changes the nature of the condensate.

Detailed properties of the PEP can be calculated explicitly. 
Interestingly, the PEP has the same stationary state 
as that of the ZRP with the hopping rate~\cite{Angel05,Schwarzkopf08}
\begin{equation}\label{zrp_rate}
u_{ZRP}(m) = 1+\frac{b}{m} + c\frac{m}{N} .
\end{equation}
The ZRP has the factorized stationary state with $f_{ZRP}(m) = \prod_{n=1}^m
[u_{ZRP}(n)]^{-1}$. In the large $N$ limit, $f_{ZRP}(m)$ becomes equal to 
$f(m)$ with $c=\lambda/\rho$. The nature of the condensed phase is
documented well in Ref.~\cite{Schwarzkopf08}. 
Hence, we make use of the results of
Ref.~\cite{Schwarzkopf08} to present the property of the PEP.

The PEP undergoes a condensation transition at the phase boundary 
$\rho=\rho_c(b) = {1}/{(b-2)}$. 
When $\rho<\rho_c$, the system is in a fluid phase with 
$p(m) \sim m^{-b} \exp(-m/m_0)$. When $\rho>\rho_c$,
the system is in a condensed phase. Here, the distribution function decays
algebraically as $p(m) \sim m^{-b}$ with an additional peak for condensates
centered at 
\begin{equation}\label{mass_c}
m_{\rm con} \sim (N \ln N)^{1/2}   .
\end{equation}
The peak height scales as $p(m_{\rm con})\sim N^{-1} (\ln N)^{-1/2}$
and the width as $\Delta m \sim N^{1/2}$. So the number of
condensates $N_{\rm con} \simeq N p(m_{\rm con}) \Delta m_{\rm con}$ 
scales as
\begin{equation}\label{number_c}
N_{\rm con} \sim (N/\ln N)^{1/2} \ .
\end{equation}
The effect of the PE is evident. It breaks a macroscopic
condensate into the number $N_{\rm con}$ of mesoscopic condensates of size
$m_{\rm con}$. The total mass of condensates is still
macroscopic, i.e., $N_{\rm con} m_{\rm con} \sim N^1$.

The PE also plays an important role in the critical scaling at
$\rho=\rho_c$, where the mass distribution is given by 
$p(m) \sim f(m)$.  The maximum value of $m$, denoted as $m_c(N$), will 
scales with $N$.  We will derive the scaling law for the maximum mass 
by using the extreme value statistics~\cite{Majumdar05_Evans06,Evans08}.
Let $F(m_c)$ be the probability that the maximum mass is $m_c$.
It is given by 
$$
F(m_c) = a^{-1} N f(m_c) \left(a^{-1} \sum_{m'=0}^{m_c} f(m') \right)^{N-1} \ ,
$$
where $a=\sum_{m=0}^\infty f(m)$ is a normalization constant.
For large $N$ and $m_c$, it can be
written as $F(m_c) \sim N f(m_c) \exp[-NQ(m_c)/a]$ where
$ Q(m_c) = \int_{m_c}^\infty dm' f(m').  $
Hence, the maximum mass $m_c(N)$ can be 
determined from the relation $Q(m_c) \sim 1/N$.
The integral with $f(m)$ in Eq.~(\ref{f_factor}) yields that
\begin{equation}
Q(m_c) = \frac{1}{2} \left( \frac{
\lambda}{2\rho N}\right)^{(b-1)/2} \Gamma\left( \frac{1-b}{2}, \frac{ \lambda
m_c^2}{2\rho N}\right) , 
\end{equation}
where $\Gamma(s,z) \equiv \int_z^\infty t^{s-1} e^{-t} dt$ is the incomplete
gamma function~\cite{Abramowitz}. It has the limiting behavior of
$\Gamma(s,z\rightarrow \infty) \simeq z^{s-1} e^{-z}$ and
$\Gamma(s,z\rightarrow 0) \simeq \Gamma(s) - z^2/s$. 
Consequently, the
function $Q(m)$ behaves as $Q(m) \simeq \frac{1}{2}  \left( \frac{
\lambda}{2\rho N}\right)^{-1} m^{-(1+b)/2} e^{ -\lambda m^2 / (2 \rho N)}$
for $(\lambda m^2) \gg N$ and  $Q(m) \sim m^{1-b}$ for 
$(\lambda m^2) \lesssim N$. Solving $Q(m_c) = 1/N$ self-consistently, we
obtain that
\begin{equation}\label{m_c_result}
m_c(N) \sim \left\{
\begin{array}{ccc}
 ( N \ln N )^{1/2} &\mbox{for}&  b < 3 , \\ [2mm]
 N^{1/(b-1)} &\mbox{for}& b \ge 3 .
\end{array}\right.
\end{equation}
This is contrasted to the ZRP without PE~($\lambda=0$), where
$m_c \sim N^{1/(b-1)}$ for all values of $b>2$~\cite{Evans08}. 
The results of Eqs.~(\ref{mass_c}) and (\ref{m_c_result}) justify 
the approximation ignoring $\mathcal{O}((m/M)^2)$ term in 
$\tilde{v}(m,m')$.

We have performed a numerical analysis to confirm our analytic results
in one-dimensional rings and  complete graphs with unbiased hopping.
The single-site mass distribution function $p(m)$ has been measured 
from Monte Carlo simulations. We have also evaluated $p(m)$ numerically by 
enumerating the analytic expressions in Eqs.~(\ref{f_prod}) and
(\ref{pm_def}).
The normalization factor $Z(N,M)$ can be evaluated
from the recursion relation $Z(N,M) = \sum_{m=0}^M f(m) Z(N-1,M-m)$.

\begin{figure}[t]
\includegraphics*[width=\columnwidth]{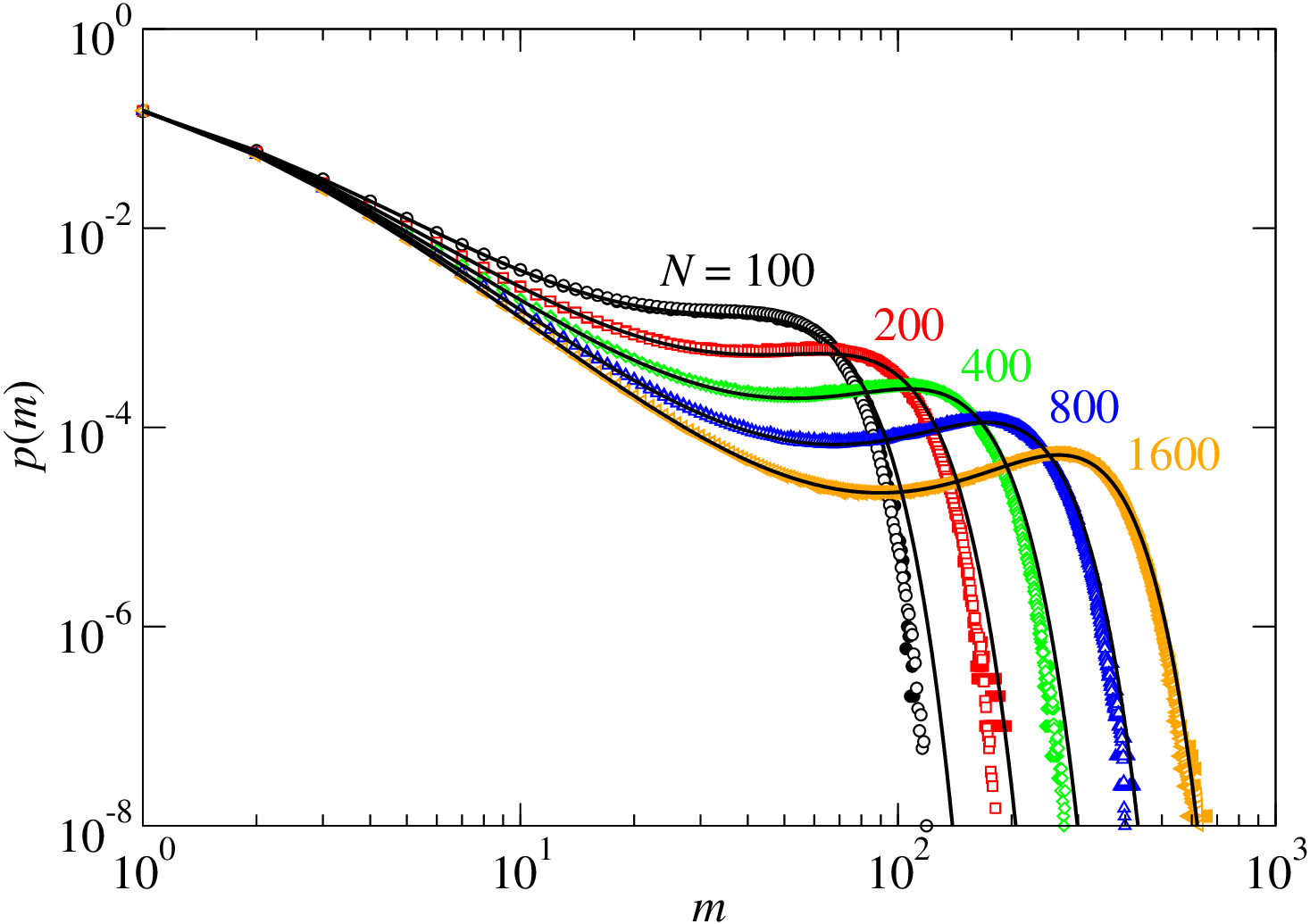}
\caption{(Color online) Mass distribution function $p(m)$ measured in
one-dimensional rings~[filled symbols] and complete graphs~[open symbols]
with $b=4$ and $\rho=4$.
The solid lines represent the curves for $p(m)$ obtained from
Eq.~(\ref{pm_def}).}\label{fig1}
\end{figure}

We present numerical data for $p(m)$ at $b=4$ and $\rho=4$ 
in Fig.~\ref{fig1}. The plots for one-dimensional rings overlap perfectly 
with those for complete graphs. This suggests that it is likely that 
the particle species distribution is indeed random even in one-dimensional 
rings.
There is a little deviation between the simulation data and the analytic
results drawn with solid lines. The deviation comes from the approximation 
that $\mathcal{O}((m/N)^2)$ term is neglected in $\tilde{v}$. 
Since $m$ is limited by the cutoff $m_{PE}\sim
N^{1/2}$, the approximation is expected to be better as $N$ increases. 
Numerical data shows that the deviation is already negligible at $N=1600$,
which justifies the approximation.

\begin{figure}[t]
\begin{tabular}{cc}
\includegraphics*[width=0.5\columnwidth]{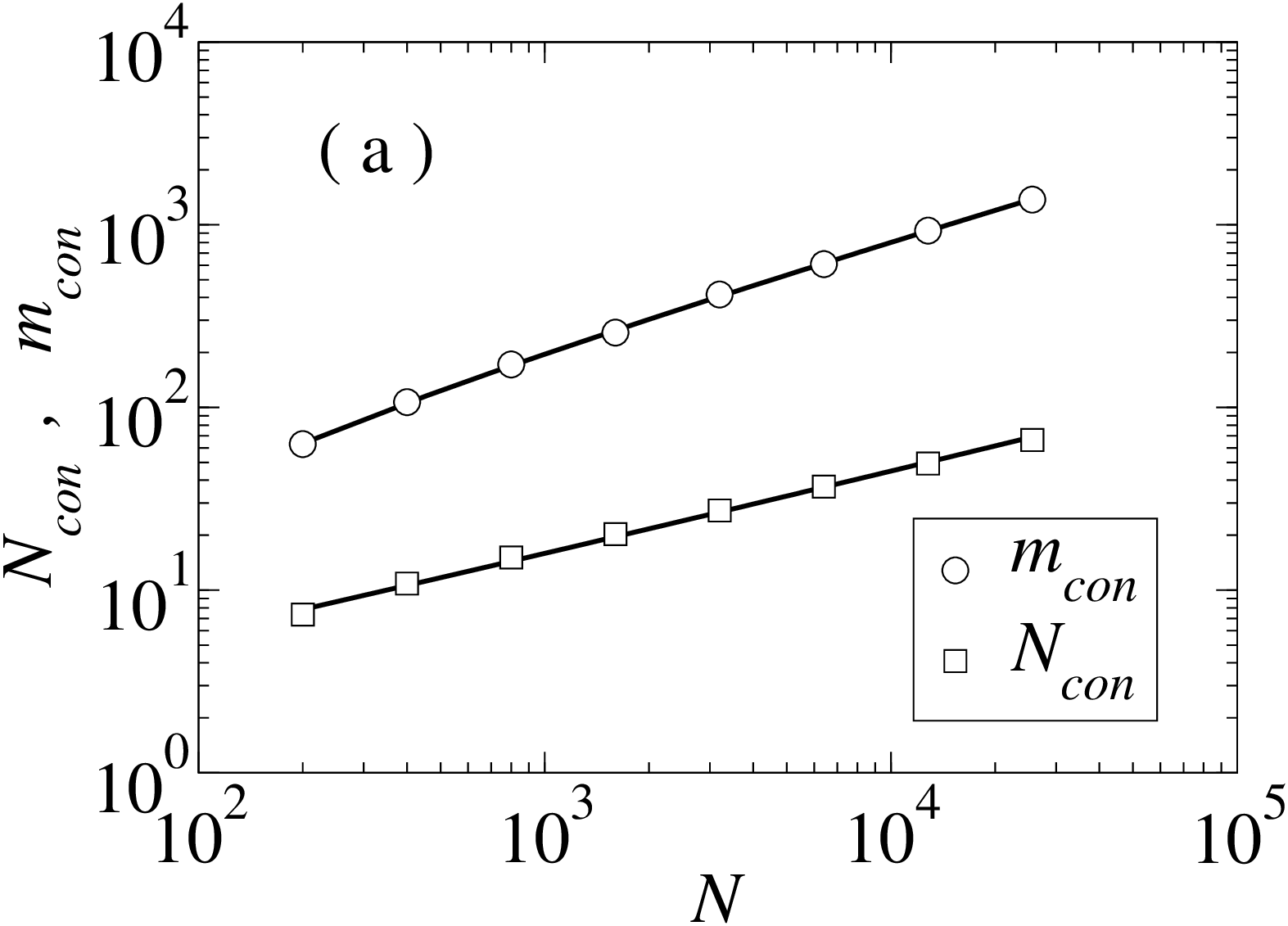} &
\includegraphics*[width=0.5\columnwidth]{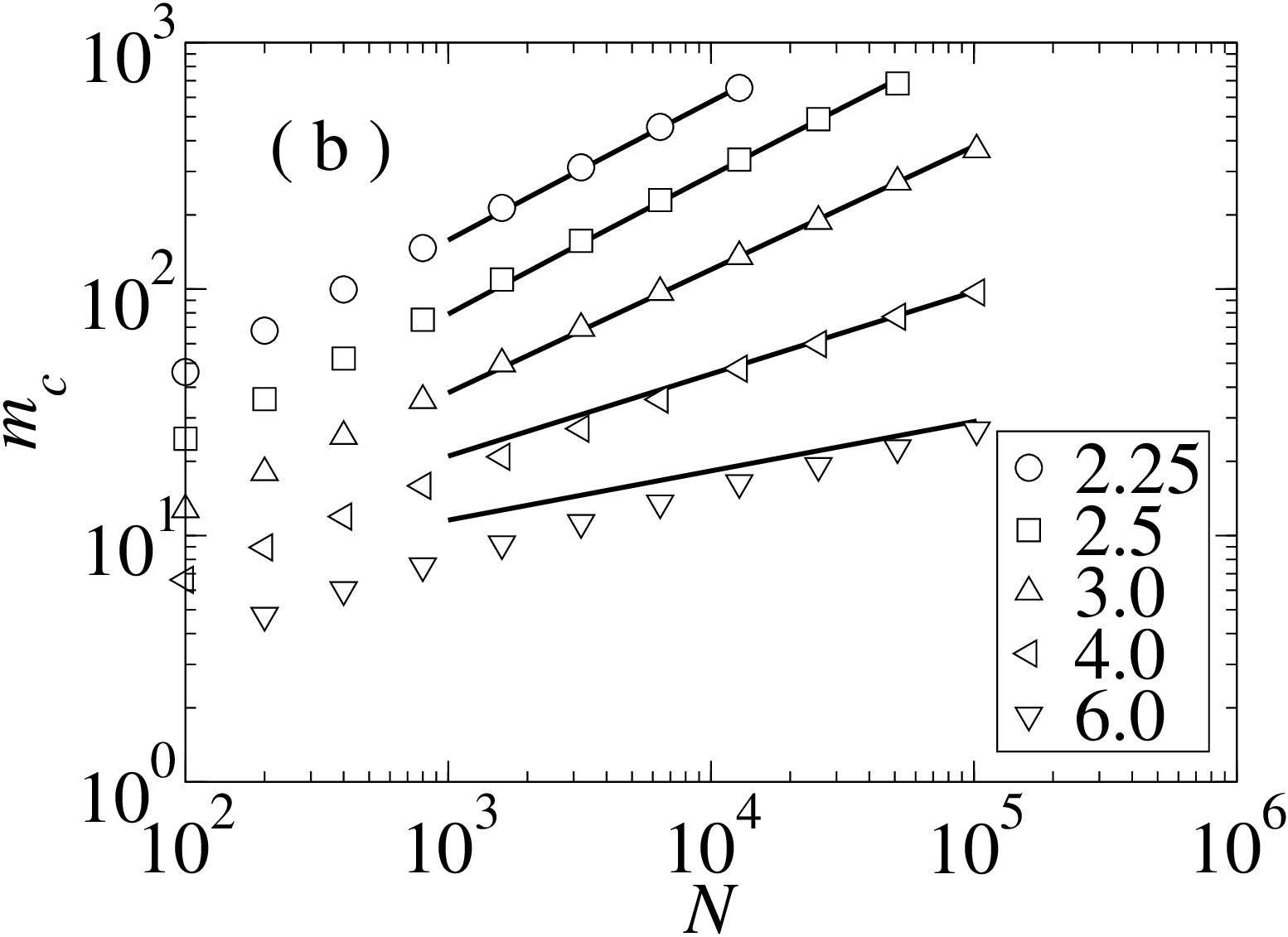}
\end{tabular}
\caption{(a) $m_{\rm con}$ and $N_{\rm con}$ in a condensed phase 
at $b=4$ and $\rho=4$. 
The fitting curves explained in the text are drawn with solid lines.
(b) The maximum mass $m_c(N)$ at the critical points $\rho = 1/(b-2)$ at
several values of $b$. Solid lines represent the analytic prediction of
Eq.~(\ref{m_c_result}).}
\label{fig2}
\end{figure}

The distribution function $p(m)$ in Fig.~\ref{fig1}
displays a broad peak representing condensates in the large $m$
region. The typical size $m_{\rm con}$
of condensates is given by the peak position, and the number $N_{\rm con}$
of condensates is given by $N_{\rm con} = N \Omega$ where $\Omega$ is the 
spectral weight of the peak. 
The spectral weight 
can be estimated as $\Omega=\sum_{m\ge m_{min}} p(m)$ where $m_{min}$ 
is the location of the local minimum of $p(m)$.
The numerical data thus obtained at $b=4$ and $\rho=4$
are presented in Fig.~\ref{fig2}(a).
In order to compare with the analytic result of Eqs.~(\ref{mass_c}) and 
(\ref{number_c}), we fitted $m_{\rm con}(N)$ and $N_{\rm con}(N)$ to the
functions $a_1 (N \ln (N/a_2))^{1/2}$ and $a_3 (N / \ln(N/a_4))^{1/2}$,
respectively. Numerical data are in good agreement with the fitting curves
drawn with solid lines for large $N$.

We have also examined the scaling behavior of the maximum mass $m_c(N)$ 
along the critical line $\rho = \rho_c = 1/(b-2)$. The
analytic theory based on the extreme value statistics predicts the scaling
law in Eq.~(\ref{m_c_result}). 
We have measured $m_c(N)$ at several values of $b$
and present them in Fig.~\ref{fig2}(b). 
The solid lines represent the analytic result of Eq.~(\ref{m_c_result}).
Numerical data are in agreement with the analytic prediction for large
values of $N$. 

In summary, we have shown that the pair exclusion can break a
macroscopic condensate into multiple mesoscopic condensates. The pair
exclusion is a weak interaction in that a particle interacts with only a
single one. Such a weak interaction is negligible in a fluid
phase. However, it becomes important during the course of
condensation and makes a macroscopic condensate fragile.
The mesoscopic condensation occurs in broader cases.
Consider a pair repulsion that a pair occupation is rejected 
with probability $\lambda<1$. Such a system is also described by the same 
product state with Eq.~(\ref{f_prod}). The cutoff $m_{PE}$
scales with the same exponent, hence the system displays 
the mesoscopic condensation for any nonzero values of $\lambda$.

The pair exclusion type interaction can play an important role in 
a broad range of phenomena. It is manifested as the self-loop constraint
in network theory, whose effects have been neglected in most
studies. Percolation is a macroscopic condensation of sites into a single 
cluster. It will be interesting to study how the pair exclusion type 
interaction affects the nature of the percolation transition. 
In fact, recent studies on the Achlioptas process~\cite{Achlioptas09} 
reveal that a slight modification in a cluster growth rule changes a 
continuous percolation transition into a discontinuous 
one~\cite{Achlioptas09,Ziff09,Cho09}. 
Another interesting application is an opinion dynamics modeled by a voter
model~\cite{votor}. It may be interesting to study the effect of hostility
between individuals on consensus formation. Hopefully our work will initiate 
further studies in those directions.

This work was supported by KOSEF
grant Acceleration Research (CNRC)~(Grant No. R17-2007-073-01001-0) and 
2007 Hongik University Research Fund.

\end{document}